\documentclass[aip,apl,preprint]{revtex4-1} 

\usepackage{graphicx}
\usepackage{textcomp}
\usepackage{hyperref}

\draft 

\begin{document}



\title{Metallic nanorings for broadband, enhanced extraction of light from solid-state emitters} 


\author{Oliver J. Trojak}
\email{o.trojak@soton.ac.uk}
\affiliation{Department of Physics and Astronomy, University of Southampton, Southampton, SO17 1BJ, United Kingdom}

\author{Suk In Park}
\affiliation{Center for Opto-Electronic Materials and Devices Research, Korea Institute of Science and Technology, Seoul 136-791, South Korea}

\author{Jin Dong Song}
\affiliation{Center for Opto-Electronic Materials and Devices Research, Korea Institute of Science and Technology, Seoul 136-791, South Korea}

\author{Luca Sapienza}
\email{l.sapienza@soton.ac.uk}\homepage{www.quantum.soton.ac.uk}
\affiliation{Department of Physics and Astronomy, University of Southampton, Southampton, SO17 1BJ, United Kingdom}



\date{\today}

\begin{abstract}
We report on the increased extraction of light emitted by solid-state sources embedded within high refractive index materials. This is achieved by making use of a local lensing effect by sub-micron metallic rings deposited on the sample surface and centered around single emitters.  We show enhancements in the intensity of the light emitted by InAs/GaAs single quantum dot lines into free space as high as a factor 20. Such a device is intrinsically broadband and therefore compatible with any kind of solid-state light source. We foresee the fabrication of metallic rings via scalable techniques, like nano-imprint, and their implementation to improve the emission of classical and quantum light from solid-state sources. Furthermore, while increasing the brightness of the devices, the metallic rings can also act as top contacts for the local application of electric fields for carrier injection or wavelength tuning. 
\end{abstract}


\maketitle 


Extracting light into free space is one of the challenges to face when dealing with solid-state emitters embedded within high-index materials. At the air interface total internal reflection can trap most of the light within the higher index material, thus preventing efficient light extraction, that can be as low as a few percent. Such an issue needs to be faced when dealing with emitters like light-emitting diodes \cite{LED} and lasers \cite{lasers} based, for instance, on quantum wells or quantum dots. In the same way as for classical light emitters, extraction efficiency has been the focus of intensive research when dealing with intrinsically dimmer sources like single-photon emitters for fundamental science \cite{Igor, Gazzano} and quantum information technology applications \cite{quantum_tech}. Amongst solid-state quantum light sources, molecular beam epitaxial quantum dots (QDs) are of particular interest as they are directly grown on a semiconductor chip (thus allowing easy integration within optical circuits), they can have lifetime-limited emission lines \cite{Richard} and can emit pure and indistinguishable single photons \cite{Pascale}. Several approaches have been followed to increase the extraction efficiency of light emitted by QDs into free space. For instance, optical cavities embedding single emitters have been fabricated to channel the emitted light into specific optical modes. Examples are micropillars, based on distributed Bragg reflectors \cite{Pascale}, nanowires \cite{Claudon} and circular grating cavities on suspended membranes \cite{Marcelo, IEEE}. Such optical cavities require the coupling of the emission from a source into a specific optical mode that the emitter needs to be resonant with. Fabrication processes can require multilayer growth (for micropillars) and deep etching (for micropillars and nanowires) or a calibrated etch (for circular grating cavities). High aspect ratio devices or suspended membranes also require non-trivial fabrication processes if one wants to include electrical tuning or injection to improve the device performances \cite{Pascale, Niels}.
To avoid complex fabrication processes and potential degradation of the emitter's optical performances due to the proximity to etched surfaces, solid immersion lenses (SILs) can be used to focus the excitation and emission light from bulk emitters and achieve enhanced free-space collection.  Macroscopic SILs (typically of $\sim$1\,mm base diameter) can be deposited onto the surface of the sample and can provide about one order of magnitude enhancement of the collection efficiency \cite{SILs}. However, given the large size of the SILs with respect to the emitters, typically, they are deposited on the surface of a sample containing a relatively high density of QDs and, after careful alignment, the light from those emitters better aligned with the SIL apex is collected. Alternatively, SILs deterministically  fabricated around a specific QD have been recently demonstrated: in situ lithography is used to create a SIL in correspondence to a single QD placed above a distributed Bragg reflector: such a device showed modest enhancements in the collection efficiency of up to a factor two \cite{Sartison}.

Here we report on a different approach to increase the extraction efficiency from single-photon emitters: we make use of the focusing properties of metallic rings deposited on the surface of the sample and centered around single quantum dots. 
Such metallic rings provide a lensing effect that concentrates the excitation light and focuses the emission vertically for enhanced free-space collection.  Compared to optical cavities, metallic rings are intrinsically broadband; they do not rely on a specifically prepared substrate and can therefore be applied to any emitter/substrate combination; they are easy to fabricate, allowing scalability and a high yield; and they can provide more than one order of magnitude increase in the light collection efficiency.



The sample under study consists of a GaAs wafer embedding a single layer of low density ($\sim$1000 \textmu m$^{-2}$) InAs quantum dots grown by molecular beam epitaxy and capped by a 95\,nm GaAs layer.  A 1\,\textmu m-thick Al$_{0.7}$Ga$_{0.3}$As sacrificial layer is present 95\,nm underneath the QD layer to allow the possible realisation of suspended membranes.


Finite-Difference Time-Domain (FDTD) simulations were performed to simulate the lensing effect of metallic rings deposited on the sample surface and to optimise their dimensions.  The QD was modelled as a dipole emitter with a Gaussian emission spectrum, placed 95\,nm below the sample surface.  The rings were designed to be composed of a 7\,nm Cr adhesion layer capped by a variable thickness ($t$) of Au.  The inner ($r$) and outer radius ($R$) of the rings were left as free parameters in the particle swarm optimization routine\cite{Kennedy} used to modify the ring dimensions in order to maximize the far-field emission through a flux-plane placed 2.5 \textmu m above the substrate surface.  The optimal design dimensions for the rings were found to be: $r$ = 220\,nm, $R$ = 540\,nm, $t$ = 60\,nm.

\begin{figure}[h!]
	\includegraphics[width=\columnwidth]{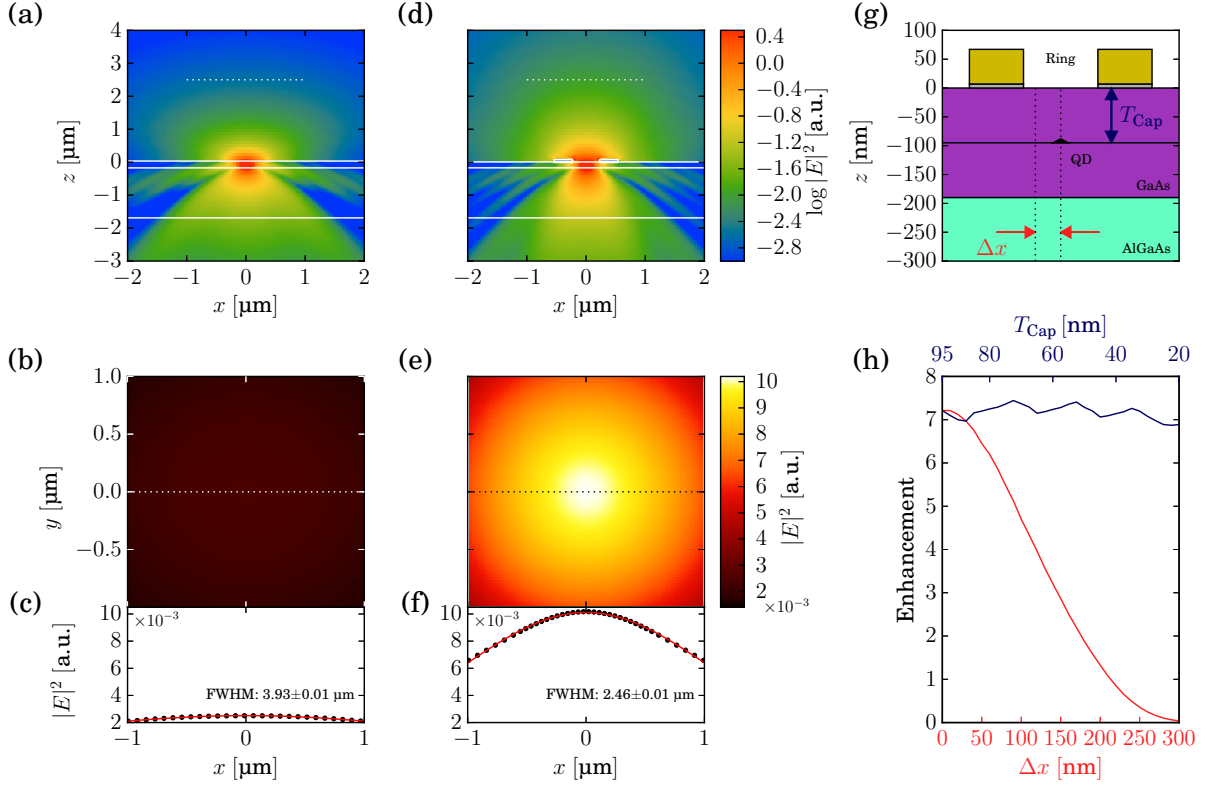}
	\caption{\label{fig1}(a) Finite-Difference Time-Domain (FDTD) simulation of the electric field of a dipole emitter placed 95\,nm below the surface.  The different material layers (from bottom to top: GaAs/Al$_{0.7}$Ga$_{0.3}$As/InAs/GaAs/air) are separated by solid white lines.  (b) Colour plot of the squared electric field profile, measured at a distance of 2.5\,$\mu$m from the substrate surface, in correspondence to the dotted line shown in panel (a). (c) Linecut of the far-field profile (symbols), in correspondence to the dotted line in panel (b), and its Gaussian fit (red line). The Full-Width-Half-Maximum (FWHM) obtained from the fit is indicated in the graph.  (d, e, f) Same as panels (a, b, c) in the presence of a Cr/Au (7\,nm/100\,nm) ring on the sample surface, centered around the emitting dipole. (g) Schematic of the sample (vertical section, not to scale). $\Delta x$ represents the horizontal position of the emitter with respect to the center of the ring and $T_{\text{Cap}}$ the vertical distance of the dipole from the sample surface.  (h) Dependence of the emission enhancement factor as a function of the position of the dipole with respect to the center of the ring ($\Delta x$, shown in red) and, with the dipole positioned in correspondence to the center of the ring, as a function of the distance of the dipole from the top GaAs surface ($T_{\text{Cap}}$, shown in blue).}
\end{figure}

In the simulations, we calculate the electric field intensity from an emitter in bulk (Fig.\,\ref{fig1}a) and measure its intensity on a plane parallel to the substrate (Fig.\,\ref{fig1}b).  In Fig.\,\ref{fig1}c,  we show a line-cut across the center of this monitor surface and fit the far-field profile with a Gaussian function that allows to evaluate the field intensity and width. We observe a far-field profile with a full-width half-maximum (FWHM) of 3.9\,\textmu m.  The same simulations are carried out when a metallic ring of optimised dimensions is placed on the sample surface, centered around the emitting dipole. As shown in Fig.\,\ref{fig1}d,e,f, the electric field is focussed by the ring: the far-field emission is brighter and its line-cut shows that the emission remains Gaussian with a FWHM reduced to 2.5\,\textmu m, thus proving the focusing of the emitted light.  By comparing the integrated intensity of the electic field over the far-field surface with and without the ring, we observe an emission enhancement of about $\times7.2$.

In Fig.\,\ref{fig1}g,h we show the effect on the enhancement factor on the horizontal displacement of the QD with respect to the center of the ring ($\Delta x$) and on the vertical displacement, obtained by varying the GaAs capping layer thickness ($T_{\text{Cap}}$). The simulation shows that, for displacements of the emitter larger than $25\,\text{nm}$ from the center, a gradual decline in device performances is observed, with a halving of the enhancement in the emission intensity for a displacement of 120\,nm (see Fig.\,1h). When varying the distance between the emitter and the surface, for a given ring design, we see that the enhancement does not vary significantly (see Fig.\,1h). This proves that geometries with emitters (like colloidal quantum dots, defects centers in diamond, fluorescent molecules) deposited on the sample surface would still benefit from the proposed ring design.



\begin{figure}[h!]
 	\includegraphics[width=0.8\columnwidth]{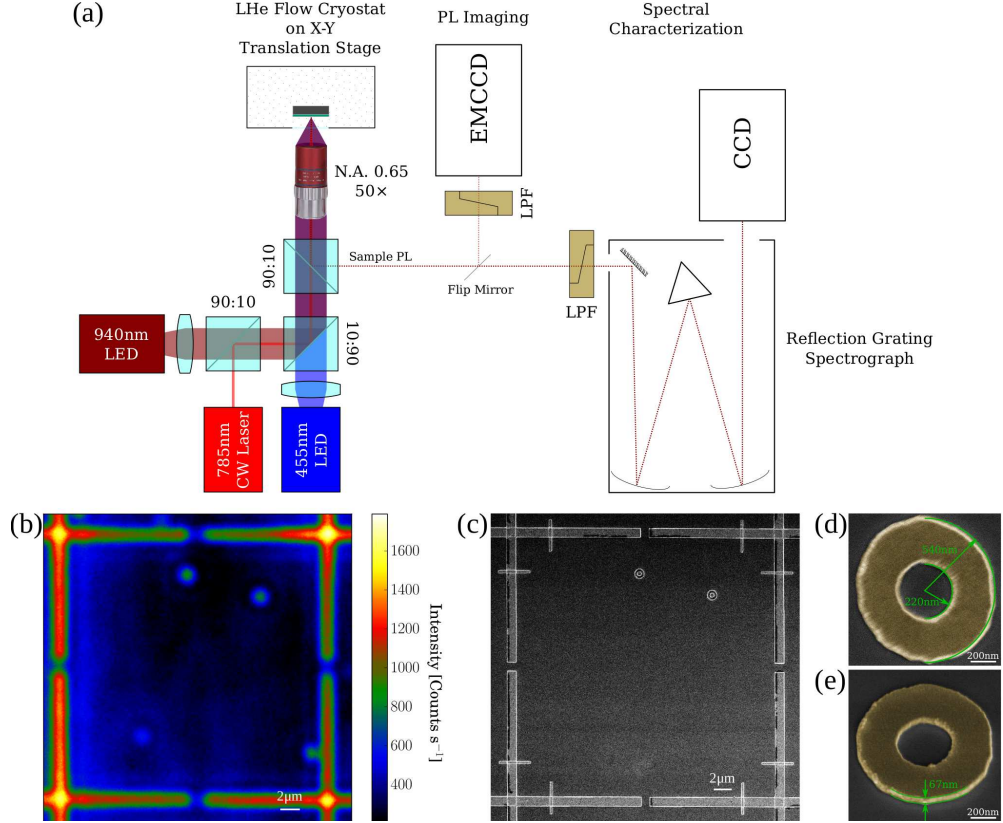}
 	\caption{\label{fig2}(a) Schematic of the confocal micro-photoluminescence (PL) set-up, composed of: a 785\,nm continuous wave (CW) laser for single QD above-band excitation; a light-emitting diode (LED) with emission wavelength centered around 940\,nm used for illumination of the sample; an LED with emission wavelength centered around 455\,nm for the wide-field excitation of the QDs. The excitation and emitted light are focussed and  collected by an objective with numerical aperture (N.A.) 0.65. The sample is mounted inside a liquid helium flow cryostat sitting on an X-Y translation stage.  The images are taken using a electron-multiplying charge-coupled device (EMCCD), preceded by long-pass-filters (LPFs) used to remove the excitation and wetting layer emission. By flipping a mirror in the optical path, the PL can be sent into a grating spectrometer for spectral characterization. (b) Example of an image, obtained under double LED illumination, showing metal alignment markers and the PL emission from several QDs (EMCCD settings: 1\,s integration, 360 acquisitions, 50$\times$ gain). (c) Scanning electron microscope (SEM) image of the metallic markers and the metallic rings fabricated around selected QDs.  (d) False-colour SEM image of a ring indicating its measured dimensions. (e) SEM image of the same ring as in panel (d) imaged with a 35\textdegree\ tilt.}
 \end{figure}

As a first step, metallic markers needed for the location of single quantum dots, are written by means of electron-beam lithography on the sample surface and metallized by thermal evaporation (Cr/Au) followed by a chemical lift off. The samples are placed on the cold finger of a liquid helium flow cryostat and cooled down to temperatures of $\sim$10\,K and characterized using a confocal micro-photoluminescence (PL) set-up, schematically shown in Fig.\,\ref{fig2}a. The same objective lens, with a numerical aperture (NA) of 0.65, is used to focus the excitation light and collect the emission.  A light-emitting diode (LED), with emission centered around 455\,nm, is used to provide wide-area above-band excitation of the quantum dots; an LED, with emission centered around 940\,nm, is used to image the metallic markers deposited on the sample surface.  The reflected light and quantum dot emission are imaged on an electron-multiplying charge-coupled device (EMCCD), after the excitation light is removed by long-pass filters (LPF). We use a PL imaging technique \cite{NComm, Jin} to locate single quantum dots with nanometer-scale accuracy and to selectively address them with a 785\,nm continuous-wave (CW) laser for spectral characterization, via a reflection grating spectrograph equipped with a CCD.


An example of the images collected on the EMCCD under double LED illumination is shown in Fig.\,\ref{fig2}b: the emission from four QDs appears as bright circular dots and the alignment marks are clearly visible.  Line-cuts from the image are fitted with Gaussian functions to determine the locations of the QDs with respect to the alignment marks: from the fits, we evaluate the errors in the PL positioning procedure to be of about 25\,nm. The QD coordinates with respect to the metallic markers are then used to perform an aligned electron-beam lithography to write rings centered around specific quantum dots, as shown in Fig.\,\ref{fig2}c. Fig.\,\ref{fig2}d,e show scanning electron micrographs of the fabricated metallic rings and their critical dimensions.



\begin{figure}[h!]
 	\includegraphics[width=\columnwidth]{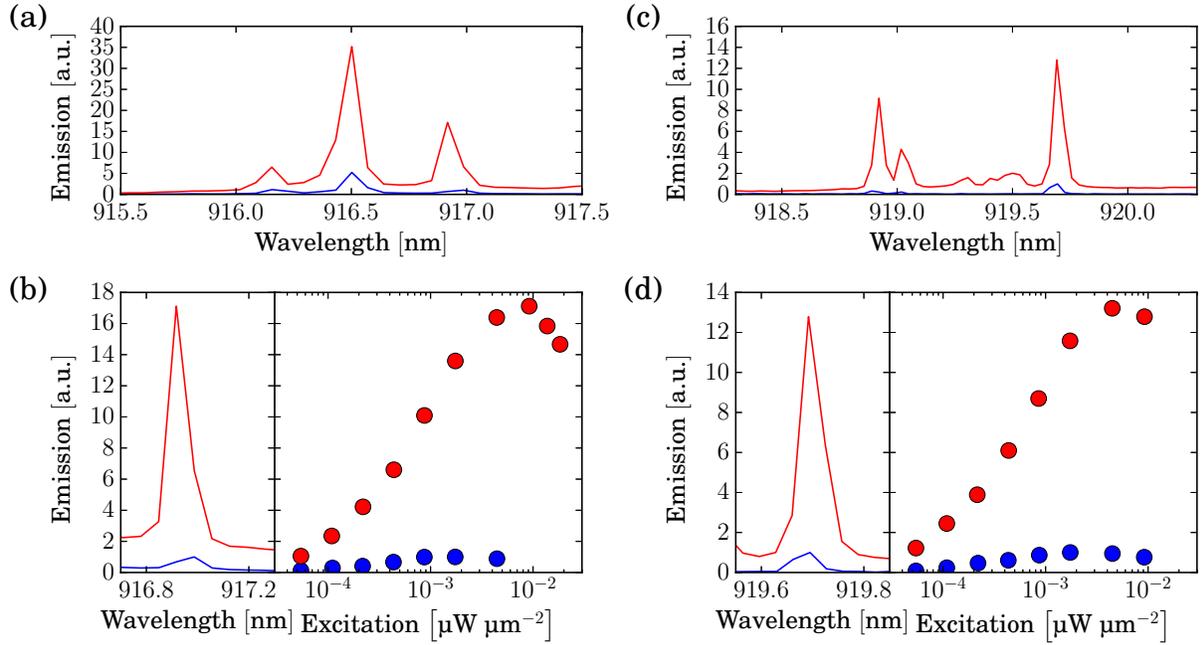}
 	\caption{\label{fig3} (a, c) Photoluminescence spectra from a single QD collected under 785\,nm CW laser excitation at QD saturation power, at a temperature of 10\,K, before (blue) and after (red) metallic ring deposition. (b, d) Left panels: Zoom-in of a photoluminescence emission line from a single quantum dot, collected at saturation level before (blue) and after (red) ring deposition.  Right panels: Emission intensities measured as a function of excitation power, for the emission lines shown in the left panels (the same colour coding is used).  The intensities are normalised to the saturation level of the lowest intensity peak from the QD in bulk.}
 \end{figure}

Example photoluminescence spectra of two QDs before and after ring deposition are shown in Fig.\,\ref{fig3}. To reliably compare intensity measurements before and after ring fabrication, we need to eliminate potential errors due to different alignement of the optics in the photoluminescence set up between different measurement runs. To this end, the intensity of the light emitted by selected bulk QDs was collected as a function of pump power before every set of measurements and used for our intensity calibration. If the calibration QDs intensities were different, correction factors to appropriately scale the recorded emission, taking into account the different alignement of the setup, were introduced. We would like to stress that such adjustments were of a maximum of $\sim$30$\%$ (therefore much lower than the reported enhancement factors) and, in order to be sure of the validity of the reported results, all the data shown below only comprise measurements where the intensity scalings reduce the emission enhancement factors. 
A zoom-in of selected quantum dot lines collected under saturation excitation power before and after ring deposition are shown in the left panels of Fig.\,\ref{fig3}b,d. The right panels show the intensity of the same lines and the enhancement of the emission as a function of laser excitation power. The QD-ring devices in Fig.\,3 show an increase in the collected intensity of about $\times17.2$ and $\times13.2$ at saturation. A different enhancement factor can be attributed to the different position of the QDs with respect to the center of the ring, as discussed in Fig.\,\ref{fig1}h. Non-ideal positions of the rings can be due to the combined error in the positioning of the emitter with respect to the alignment marks and the error introduced in the aligned electron-beam lithography step. Our fabrication process has proved to be reliable, with 76$\%$ of more than 30 fabricated devices showing enhancements in the emitted intensities. Failures are attributed to incorrect ring positions, validated by scanning electron microscopy analysis. 

To rule out the role of possible plasmonic effects in the modification of the emission properties of the quantum dots under study, we have carried out time-resolved measurements of the spontaneous emission and observed no modifications in the lifetimes within the measurement error. Plasmonic effects are unlikely in our devices, given the distance of the emitter from the metallic rings (about 100\,nm) and the relatively large size of the rings (inner diameter of 440\,nm). However, the fabrication technique here presented are fully compatible with the realisation of a deterministic coupling between single emitter and plasmonic fields \cite{plasmon}.
 
It is worth noting that different QD emission lines are enhanced by different amounts by the rings: this can be a result of the modification in the excitation spot due to the focusing of the excitation laser light by the ring. A modified excitation spot can excite differently the various QD charged states, resulting in different intensities for different emission lines, as already reported and discussed for instance in Ref.\cite{Hartmann, Sartison}. A better understanding of this effect would be provided by the implementation of charge-tunable devices that allow to address specific charged states in the quantum dots \cite{charge_tune1, charge_tune2}. We would like to stress that our metallic rings are fully compatible with charge-tunable devices, where the ring itself, while enhancing the extraction efficiency, could also be used as a top contact to apply the electric field.

From the simulations shown in Fig.\,1, an enhancement factor of about 7 on the total emission from a test dipole was predicted. In our measurements, we are considering the enhancement of single QD lines, since they are responsible for single-photon emission, relevant for quantum information technology applications. To compare our results to the FDTD simulations, we integrate the emission intensity from single QDs over a 20\,nm wavelength window (a range broad enough to cover the QD single lines emission) before and after ring deposition. From the integrated intensities, we measure an average enhancement factor of about a factor 4, an experimental value compatible with the results of our FDTD simulations.

In conclusion, we have demonstrated that metallic rings can be used to increase the brightness of single quantum dot lines, up to a factor of 20. Such an enhancement is intrinsically broadband, since it is the result of a lensing effect on the excitation and collected light, and does not rely on the coupling to an optical cavity.  Therefore, metallic rings could be used to improve the brightness of any kind of solid-state device.  Additionally the rings can be integrated for instance with bottom mirrors and polymer SILs \cite{Robert} to further enhance the collected intensity. Such metallic rings act as local lenses deposited around specific emitters and could be implemented as top contacts for charge-tunable devices \cite{Richard} (that often require SILs to increase the extraction efficiencies), for wavelength tuning via Stark effect \cite{Stark} or for charge-noise reduction via electric fields \cite{Pascale}. Overall they are relatively easy to fabricate with high yield and could allow scalability, given that the dimensions in play are compatibile with photolithography and nano-imprint techniques.

\begin{acknowledgments}
We would like to thank Barnaby Sleat and Jason Tam for their early contributions to this work and Zondy Webber for his help in the lithographic process. 
SIP and JDS acknowledge the support from the KIST institutional program of flagship.
\end{acknowledgments}

\bibliographystyle{aipnum4-1}

\end{document}